\documentclass[a4paper,fleqn]{cas-dc}

\usepackage[numbers]{natbib}

\usepackage{tikz}
\usetikzlibrary{positioning,arrows.meta,shapes,fit,backgrounds}

\usepackage{helvet} 

\usepackage{amsmath}
\usepackage{amssymb}
\usepackage[ruled,vlined,linesnumbered]{algorithm2e}

\begin{document}
\let\WriteBookmarks\relax
\def\floatpagepagefraction{1}
\def\textpagefraction{.001}

\shorttitle{Spectral Biclustering-Driven Scalability for Post-Hoc Explainability in Recommender Systems}
\shortauthors{J.L. Salmeron and I. Ar\'evalo}

\title[mode=title]{Spectral Biclustering-Driven Scalability for Post-Hoc Explainability in Recommender Systems}

\author[1]{Jose L. {Salmeron}}
[
orcid=0000-0001-7811-3716]
\ead{joseluis.salmeron@cunef.edu}
\ead[url]{joselsalmeron.github.io}
\affiliation[1]{organization={School of Engineering, CUNEF Universidad},
                city={Madrid},
country={Spain}}


\author[2]{Irina {Ar\'evalo}}
[
   orcid=0000-0002-8967-0208]
\cormark[1]
\ead{irina.arevalo@upm.es}
\ead[URL]{irinaarevalo.github.io}


\affiliation[2]{organization={School of Civil Engineering, Universidad Politécnica de Madrid}, 
city={Madrid},
country={Spain}}



\cortext[cor2]{Corresponding author}


\begin{abstract}
Explainability in recommender systems is essential for ensuring transparency, accountability, and trust, yet existing post-hoc methods often encounter severe scalability challenges. Observation-level deletion diagnostics offer a counterfactual way to analyze recommendations by retraining models after removing individual users or items, but their cost grows rapidly with dataset size. To improve the practical tractability of this analysis, this paper introduces a block-deletion diagnostic framework that uses spectral biclustering to group users and items and then removes entire blocks of interactions. This formulation reduces the number of retraining procedures relative to finer-grained deletion strategies and produces explanations at the level of user segments, item groups, and their interactions. The framework is evaluated on two representative recommender paradigms, Singular Value Decomposition and Neural Collaborative Filtering, using the MovieLens and Amazon datasets. The results show that top-ranked recommendations are often more sensitive to specific interaction blocks than lower-ranked ones, with some blocks acting as supporting evidence and others having a detrimental effect on recommendation quality. The analysis also indicates that user segments differ in their sensitivity to block removal, suggesting heterogeneous levels of reliance on localized interaction patterns. These findings provide diagnostic information that is not directly visible through standard recommendation metrics. Overall, the results suggest that block-deletion diagnostics offer a practical and model-agnostic post-hoc analysis framework for recommender systems, while also highlighting that the resulting explanations depend on the chosen block structure.
\end{abstract}




\maketitle

\section{Introduction}

Explainability in Artificial Intelligence (AI) has become increasingly critical due to the widespread deployment of AI systems across high-stakes domains such as healthcare, finance, and autonomous systems. Transparent and interpretable models enable stakeholders to understand the rationale behind algorithmic decisions, fostering trust and promoting accountability \cite{Saeed2023}. Moreover, explainable AI (XAI) plays a vital role in identifying biases, uncovering failure modes, and ensuring compliance with emerging regulatory frameworks, such as the European Union's AI Act \cite{eur2021ai}. Beyond ethical and legal considerations, interpretability supports practical decision-making by providing actionable insights from complex models, allowing domain experts to validate predictions and improve system reliability. As AI systems become more complex and opaque, methods that balance performance with interpretability become increasingly important for safe and responsible adoption.

There are several approaches in XAI that aim to provide interpretability from different perspectives. Intrinsic methods integrate interpretability directly into the model by using transparent architectures such as decision trees, linear regressions, or attention-based neural networks \cite{lipton2018mythos}. In contrast, post-hoc methods provide explanations for trained models without altering their structure and include feature attribution, counterfactual example generation, and sensitivity analysis \cite{Ribeiro2016}. XAI approaches can also be classified as global, aiming to understand the overall behavior of the model, or local, explaining individual predictions. This diversity allows interpretability methods to be adapted to different levels of model complexity and user needs. Even for neural networks, which are often considered black-box models, specific explainability methods have been proposed to provide insights into their internal decision-making processes \cite{guerrero.2020}.

This work introduces block-deletion diagnostics, a framework that extends observation-level deletion to the level of user and item clusters in order to improve the practical tractability of retraining-based analysis. Specifically, spectral biclustering is used to partition the interaction matrix, after which entire user--item blocks are systematically removed prior to retraining. This produces explanations at the level of user segments, item groups, and their interactions, while reducing the number of retraining procedures relative to finer-grained deletion strategies. Because the analysis is based on actual data removal and retraining, it provides a model-agnostic counterfactual diagnostic view of recommendation behavior.

The approach is evaluated on both matrix factorization (SVD) and Neural Collaborative Filtering (NCF) models, using the widely studied MovieLens and Amazon datasets. The analysis suggests that block-deletion diagnostics can reveal informative patterns of influence: top-ranked recommendations are often more sensitive to a limited number of clusters, some user segments exhibit greater dependence on localized signals, and certain blocks have a detrimental effect on recommendation quality. These effects are not directly visible through traditional performance metrics, highlighting the diagnostic value of this work for both researchers and practitioners.

The contributions of this research are three-fold:

\begin{itemize}
\item A block-deletion diagnostic framework is introduced for analyzing recommender systems through retraining-based counterfactual interventions at the level of user--item blocks.
\item The framework combines spectral biclustering with retraining-based diagnostics to study explanations at the level of user segments, item groups, and their interactions.
\item The approach is evaluated on SVD and NCF models using the MovieLens and Amazon datasets, showing that it can identify supportive and detrimental interaction blocks and reveal differences in the sensitivity of user segments.
\end{itemize}



This research does not propose a new clustering algorithm or a new attribution mechanism in isolation. Rather, its contribution lies in reformulating retraining-based explainability for recommender systems at the level of user--item blocks instead of individual observations. In this way, block-deletion provides a model-agnostic counterfactual diagnostic procedure that is more computationally tractable than finer-grained deletion strategies, without relying on surrogate models or gradient-based approximations.

Existing deletion-based diagnostics are theoretically appealing but become increasingly expensive as the number of users or items grows. This work approaches that limitation as a granularity-control problem: instead of performing deletion at the level of individual observations, the interaction matrix is organized into user--item blocks and retraining-based counterfactual analysis is carried out at the block level.

Spectral biclustering is not used here as a clustering contribution in its own right, but as an enabling mechanism for defining structured deletion units over the interaction matrix. The contribution of this work is the combination of biclustering with retraining-based diagnostics to make counterfactual analysis more computationally tractable at the block level. This integration allows the effect of groups of interactions to be studied more efficiently than finer-grained deletion strategies, while still providing informative diagnostic signals in the experimental settings considered.


The remainder of the paper is organized as follows. Section \ref{sec:related_work} reviews related works on explainability in recommender systems. Section \ref{sec:methodology} presents the methodological background needed for this research. Section \ref{sec:methodological_proposal} introduces the proposed block-deletion diagnostic framework, detailing the clustering strategy, deletion procedure, and evaluation metrics. Section \ref{sec:experiments} describes the experimental setup, including datasets, models, and baselines, and reports the results of the analysis. Finally, Section \ref{sec:conclusions} concludes the paper and outlines promising directions for future research.

\section{Related works}\label{sec:related_work}

Explainability in recommender systems has become a rapidly growing field, intersecting research in machine learning interpretability, human-computer interaction, and information retrieval. 

Below, the authors provide an overview of the most relevant lines of work, with an emphasis on their scalability and applicability to large-scale, model-agnostic settings.

\subsection{Intrinsically interpretable models}

Intrinsically interpretable models embed interpretability directly into the design of recommender models. Unlike post-hoc approaches, which attempt to explain an already trained black-box system, intrinsically interpretable models are constructed such that their outputs and intermediate representations are inherently transparent. This category includes a wide range of techniques, from linear and factorization-based models to rule-based and attention-driven architectures.

A prominent category of intrinsically interpretable models in recommender systems is founded on matrix factorization with explicit semantics. In these approaches, the latent factors learned during factorization are enriched with interpretable dimensions, such as genres, tags, or item attributes, enabling a direct mapping between latent representations and meaningful domain concepts. This augmentation facilitates not only accurate predictions but also interpretable recommendations, allowing users and system designers to understand the reasoning behind each suggested item. Such methods contrast with purely latent approaches, where the learned factors often lack semantic meaning and, consequently, transparency \cite{rendle2012factorization, zhang.14}. By aligning the learned factors with observable metadata, these models enable natural explanations in terms of user preferences over meaningful attributes (e.g., recommended because you liked action movies). Such models have been successfully applied in domains like movie and product recommendation, where item features are readily available.

Another line of research employs rule-based systems to encode recommendations as human-readable logical rules \cite{wang2018explainable}. For instance, mining association rules can generate explanations such as "users who bought X also tend to buy Y". These methods are highly interpretable by design but often suffer from sparsity and limited generalization, as the rules may fail to capture complex nonlinear interactions between users and items.

More recently, attention-based architectures have been proposed to provide intrinsic interpretability, using learned weights to highlight important items, features, or interactions \cite{cheng2018interpretable}. In sequential recommendation tasks, for example, attention weights can be visualized to indicate which past user behaviors most influenced a prediction. However, attention has been criticized as a potentially unreliable proxy for explanation, since high attention weights do not always correspond to true causal importance.

Despite their advantages, intrinsically interpretable models face notable limitations. Their reliance on transparent structures often restricts the capacity of the model, leading to lower predictive performance compared to complex neural recommenders or ensemble methods. In addition, the intrinsically interpretable models are frequently domain-dependent: while attribute-aware factorization works well in scenarios with rich side information, it is less effective in domains with sparse or unstructured data. Moreover, the trade-off between fidelity and interpretability remains a central challenge, as improving one often comes at the cost of the other.

\subsection{Post-Hoc model-agnostic explanations}

Post-hoc model-agnostic explanation methods aim to generate interpretable justifications for the predictions of arbitrary recommendation models without requiring access to their internal parameters or training processes. This flexibility allows these techniques to be applied across a wide range of recommendation algorithms, from traditional matrix factorization and nearest-neighbor methods to complex deep learning architectures.

Early approaches in this area adapted techniques from supervised learning, such as perturbation-based explanations exemplified by LIME \cite{Ribeiro2016} and SHAP \cite{lundberg2017shap}, which approximate the local decision boundary of a model using interpretable surrogate models. Gradient-based attribution methods, originally developed for deep networks in vision and natural language processing tasks \cite{sundararajan2017axiomatic}, have been adapted for neural recommender systems to produce feature-level importance scores for user–item interactions.

In the context of recommender systems, several strategies have been explored to generate user-understandable explanations. Neighborhood-based rationales \cite{zhang.14} justify recommendations by highlighting similar users or items that contributed most to a predicted rating. Counterfactual reasoning approaches \cite{deldjoo2021counterfactuals} identify minimal changes in input features or user profiles that would alter the recommendation outcome, providing actionable insights. Causal embeddings \cite{bonner2018causal} aim to disentangle spurious correlations from genuine causal effects, offering more robust explanations in the presence of confounding factors.

Despite their versatility, post-hoc model-agnostic methods often face significant scalability challenges. Perturbation-based approaches require generating and evaluating numerous perturbed inputs, resulting in high computational overhead for large-scale recommender systems. Gradient-based methods, while more efficient, may struggle with highly non-linear or sparse models, which are common in real-world platforms with millions of users and items. Recent research has focused on developing hybrid techniques that combine model-agnostic explanations with surrogate models, dimensionality reduction, or pre-computed influence scores to reduce computational costs. These advances aim to preserve interpretability while enabling deployment in industrial-scale systems where latency and throughput are critical.

\begin{table*}[ht]
\centering
\footnotesize
\caption{Comparison of post-hoc model-agnostic explanation methods}
\begin{tabular}{|l|l|l|p{3.7cm}|}
\hline
\textbf{Method} & \textbf{Explanation type} & \textbf{Strengths} & \textbf{Limitations} \\ \hline\hline
LIME \cite{Ribeiro2016} & Local, Surrogate Model & Model-agnostic, interpretable & Computationally expensive, sensitive to perturbation \\ \hline
SHAP \cite{lundberg2017shap} & Local, Additive Feature Attribution & Theoretically grounded, consistent & High computational cost for large models \\ \hline
Integrated gradients \cite{sundararajan2017axiomatic} & Local, Gradient-based Attribution & Efficient, easy to implement & Assumes linearity along the path \\ \hline
Neighborhood-based \cite{zhang.14} & Local, Similarity-based & Intuitive, user-level insights & May not generalize well to unseen data \\ \hline
Counterfactuals \cite{deldjoo2021counterfactuals} & Local, What-if Analysis & Actionable insights, user-centric & May require retraining or complex simulations \\ \hline
Causal embeddings \cite{bonner2018causal} & Local, Causal Inference & Robust to confounding, interpretable & Requires causal assumptions, complex modeling \\ \hline
\end{tabular}
\end{table*}

Post-hoc model-agnostic explanation methods offer a flexible framework for interpreting complex recommendation models, providing actionable insights while remaining model-agnostic. Their adoption in large-scale platforms requires careful consideration of computational efficiency, explanation granularity, and real-world constraints, highlighting the need for continued research on scalable and faithful explanation methods.

\subsection{Explanations modalities}

Explanations in recommender systems can be presented through various modalities, each offering distinct advantages and challenges in terms of interpretability, user engagement, and scalability. These modalities include natural language explanations, visual explanations, and hybrid approaches that integrate symbolic reasoning with neural methods.

Natural language explanations aim to provide users with personalized and human-readable justifications for recommendations. Recent advances have employed neural text generation techniques to produce such explanations. For instance, Chen et al. \cite{chen2021generate}  introduced a hierarchical sequence-to-sequence model that generates free text explanations by leveraging item features and user preferences. This approach improves user understanding and trust in the system. Similarly, Li et al. \cite{li2022personalized} proposed a personalized prompt learning framework that incorporates user and item identifiers into pre-trained language models, improving the relevance and personalization of generated explanations. However, while these methods improve user satisfaction, they also raise concerns regarding the factual correctness and computational overhead associated with generating high-quality natural language explanations.

Visual explanations utilize graphical representations to convey the rationale behind the recommendations. Techniques such as highlighting influential latent dimensions or displaying item embeddings in reduced-dimensional spaces aim to provide intuitive insights into the recommendation process. A comprehensive survey by \cite{zhang2024visualization} reviewed various visualization techniques, categorizing them according to the purpose, scope, method, and format of the explanation. These visualizations can improve transparency and user trust, especially when dealing with complex models. However, they may be less effective for highly abstract models with numerous latent factors, as the interpretability of visual representations diminishes with increased model complexity.

Hybrid explanations combine the strengths of neural networks and symbolic reasoning to offer interpretable and scalable solutions. Neuro-symbolic AI integrates the pattern recognition capabilities of neural networks with the logical reasoning of symbolic AI, addressing the limitations of each approach. This integration enhances AI systems' ability to reason and learn from data while maintaining transparency \cite{WU2023110481}. In the context of recommender systems, hybrid models can mitigate issues such as data sparsity and cold-start problems by incorporating explicit knowledge and logical rules. For example, a counterfactual reasoning framework has been proposed to identify minimal changes in user profiles that would alter recommendation outcomes, providing actionable insights for users \cite{deldjoo.2024}. These hybrid approaches show promise in balancing interpretability and performance, making them suitable for deployment in real-world applications.

\subsection{Scalability challenges}

Despite substantial progress in post-hoc explanation methods, scalability remains a central challenge in recommender systems. Many approaches rely on repeated perturbations, sampling, or retraining to approximate the underlying model's behavior, which quickly becomes infeasible in large-scale platforms with millions of users and items. The computational cost of generating local explanations for every recommendation can lead to prohibitive latency, affecting the feasibility of real-time deployment. Furthermore, sparse interaction data, high-dimensional user and item feature spaces, and dynamic updates exacerbate the difficulty of producing accurate and timely explanations.

Several strategies have been proposed to address these scalability challenges. Parallelization techniques and distributed computation can reduce runtime for perturbation-based methods, allowing simultaneous evaluation of multiple perturbed inputs \cite{abdollahi2017using}. Approximation strategies, such as surrogate models or approaches based on influence functions, attempt to capture the essential behavior of the recommendation model without exhaustive computation \cite{chen2021fast}. Although these methods improve efficiency, they often introduce trade-offs in fidelity or generalizability, raising concerns about the accuracy and reliability of the explanations produced.

Another major challenge is the combinatorial explosion associated with interaction-level explanations in large user-item matrices. Techniques that generate explanations at the user, item, or interaction level must balance explanatory granularity with computational feasibility. Sampling-based methods, dimensionality reduction, or caching of pre-computed influence scores have been explored as practical solutions. However, the deployment of faithful, real-time, and model-agnostic explanations at the industrial scale remains an open research problem, necessitating continued exploration of efficient algorithms that preserve interpretability and transparency without compromising system performance.

This research is positioned at the intersection of these research streams. Unlike intrinsic approaches, structural constraints are not imposed on the recommender. 


In contrast to existing post-hoc methods, this work focuses on improving the practical tractability of retraining-based diagnostics through block-level deletion. Specifically, the proposed framework is model-agnostic and is designed to reduce the computational burden associated with interaction-level ablations. The empirical results show that the approach is effective on the benchmark datasets considered in this study; however, its behavior at much larger scales should be interpreted as a promising direction rather than a fully established property.

\subsection{Biclustering approaches and justification}

Biclustering refers to the simultaneous clustering of the rows and columns of a matrix, producing submatrices (blocks) that exhibit coherent patterns across both dimensions, a task formalized in early work on co-clustering and block clustering \cite{Hartigan1972, Dhillon2001}. Spectral biclustering, in particular, uses a singular value decomposition of the normalized data matrix to reveal latent checkerboard structures, which has been shown to capture meaningful co-variation patterns in diverse domains \cite{Kluger2003, Dhillon2001}. This approach is especially suitable for recommender system diagnostics because the interaction matrix inherently exhibits two-mode structure (users and items), and spectral techniques directly exploit the joint geometry of both modes.

Alternative biclustering algorithms have been proposed in the literature. For example, the mean-squared residue approach introduced by Cheng and Church identifies biclusters by minimizing an empirical score, and is widely used in gene expression analysis \cite{Cheng2000}. Probabilistic co-clustering and information-theoretic methods formulate biclustering as a model‑based inference problem, assigning rows and columns to clusters based on likelihood or mutual information criteria. Other techniques based on alternating optimization or block‑diagonal structure have also been studied \cite{Ailem2017}. However, many of these methods are either designed for densely structured biological data, produce overlapping or very small blocks that are not well suited for scalable diagnostics, or introduce significant model assumptions and computational overhead that complicate integration with model‑agnostic retraining procedures.

In contrast, spectral biclustering provides a para\-meter‑light, computationally efficient, and interpretable mechanism for simultaneous row–column partitioning that aligns with the objectives of scalable and faithful post‑hoc explanation. Its reliance on matrix decompositions that can be efficiently computed even for large sparse matrices makes it particularly appropriate for recommender system interaction data, where capturing latent user-item co‑variation patterns in block form is essential for meaningful influence estimation.

\section{Methodological background}\label{sec:methodology}

\subsection{Explainable recommendations}

Explainability constitutes a fundamental requirement in recommender systems, given their cri\-tical role in the creation of large-scale content, the shaping of user experiences, and the influence of decision-making processes~\cite{arevalo.2025, zhou.2023}. Despite their utility, recommendation algorithms often inherit systematic biases originating from user behavior or intrinsic item properties~\cite{chen.2023, deldjoo.2024}. These biases can distort model outputs, amplifying existing inequities and undermining system fairness. Consequently, improving explainability is closely linked to promoting transparency, fairness, and user trust.

Bias mitigation strategies can be categorized into three principal stages~\cite{zhou.2023}:  
(a) Pre-processing, where balancing or annotation techniques reduce bias at the data source;   (b) In-process interventions, which incorporate fairness constraints or regularization during model training; and  (c) Post-processing approaches, which re-rank outputs to correct disparities~\cite{deldjoo.2024, li.2023}.  

From the perspective of explainability, the literature distinguishes between model-intrinsic and model-agnostic approaches~\cite{zhang.2020}. Model-intrinsic methods design inherently interpretable architectures, whereas model-agnostic approaches generate explanations independently of the underlying model.

Inherently interpretable techniques, such as content-based filtering~\cite{vig.09} and neighborhood-based collaborative filtering~\cite{herlocker.00, sarwar.01}, provide transparent rationales that align closely with system behavior. However, their predictive performance often deteriorates in high-dimensional or complex recommendation contexts. Conversely, latent factor and neural models, including Neural Collaborative Filtering, achieve higher accuracy at the expense of interpretability. To address this trade-off, hybrid methods attempt to align latent dimensions with external features~\cite{bauman.17, lu.18, zhang.14} or to preserve neighborhood structures~\cite{abdollahi2017using}, thereby improving interpretability without fully sacrificing predictive power.

Model-agnostic techniques, such as SHAP and LIME, approximate model behavior using local surrogate models or feature attribution mechanisms. While offering flexibility across architectures, these methods may not fully reflect the behavior of the original model in all settings \cite{Ribeiro2016, Tohidi2024}. Influence functions~\cite{Koh2017} represent another approach, estimating the contribution of individual training samples via gradient-based approximations. Although computationally efficient, they may inadequately capture the complex dynamics associated with full retraining.

This work considers deletion-based diagnostics, which involve retraining the model after selectively removing specific users or items. The resulting perturbations in recommendation outcomes can be evaluated using ranking-oriented metrics such as MAP and NDCG. In contrast to local approximation methods, this methodology provides a retraining-based way to study how changes in the training data affect recommendation behavior. Such diagnostics can support system debugging, inform data curation strategies, and facilitate robustness evaluation.

While interpretable-by-design models remain essential for user-facing transparency, deletion-based diagnostics offer a model-agnostic way to analyze how training data affect recommendation outcomes, particularly from the perspective of developers and evaluators. By examining the effect of removing specific data units on system behavior, this approach can complement accuracy-oriented evaluation and support more transparent and accountable recommendation analysis.

\subsection{Neural Collaborative Filtering}

Neural Collaborative Filtering represents a deep learning–based extension of collaborative filtering (CF), a class of methods that predict user preferences by exploiting historical interaction data. Traditional CF approaches, such as matrix factorization, model user–item interactions through the inner product of latent vectors that encode users and items as continuous feature representations. While effective in practice, this linear formulation imposes inherent limitations in capturing complex, non-linear relationships.

To overcome these restrictions, He et al.~\cite{He.2017} proposed NCF, a framework that substitutes the inner product with a neural architecture capable of learning arbitrary interaction functions directly from data. NCF employs embedding layers for both users and items, which are subsequently processed through two complementary components: (a) the Generalized Matrix Factorization (GMF), which captures linear effects via element-wise multiplication of embeddings, and (b) a Multi-Layer Perceptron (MLP), which concatenates embeddings and models non-linear dependencies through multiple hidden layers. The outputs of GMF and MLP are integrated in the NeuMF layer, followed by a sigmoid activation to generate the final prediction.

Compared to classical factorization, NCF accommodates auxiliary information such as user demographics or item metadata, thus enhancing its ability to model complex preference structures. This adaptability has enabled its application across diverse domains, including healthcare~\cite{Ponnusamy.23}, education~\cite{Mulyana_Rumaisa_2024}, tourism~\cite{marzuki.2024, wei.2025}, and real estate~\cite{venkatesh.2024}. Nevertheless, the framework is computationally demanding, particularly in large-scale recommendation settings, and its reliance on deep architectures exacerbates the interpretability problem, leaving the model as a black box to end users and practitioners.

NCF further inherits the cold-start challenge typical of CF methods, whereby recommendations for new users or items without interaction history remain unreliable. Mitigation strategies include hybridization with content-based methods, the integration of knowledge graphs, transfer learning, and active learning techniques.

In this research, NCF is employed alongside Singular Value Decomposition within the proposed deletion-based diagnostics framework. This dual evaluation illustrates the framework’s model-agnostic properties and allows for a direct comparison between deep, non-linear architectures and linear matrix factorization techniques in identifying influential users and items. The opaque nature of NCF underscores the necessity of post-hoc diagnostic tools capable of yielding interpretable insights into complex recommendation pipelines.

\subsection{Singular Value Decomposition in recommender systems}

Singular Value Decomposition (SVD) constitutes a foundational matrix factorization method that has been extensively adopted in collaborative filtering for recommender systems~\cite{koren2009matrix}. Given a real-valued user–item interaction matrix $A \in \mathbb{R}^{m\times n}$, the SVD decomposes $A$ as $A = U \Sigma V^{\top}$, where \(U \in \mathbb{R}^{m \times m}\) and \(V \in \mathbb{R}^{n \times n}\) are orthogonal matrices containing the singular vectors left and right, respectively, and \(\Sigma \in \mathbb{R}^{m \times n}\) is a diagonal matrix whose nonzero entries are singular values \(\sigma_1 \geq \sigma_2 \geq \dots \geq \sigma_r > 0\), with \(r = \mathrm{rank}(A)\).

By truncating the decomposition to the largest singular values $k$, the compact representation $A_k = U_k \Sigma_k V_k^{\top}$ provides the best rank-$k$ approximation of $A$ in both spectral and Frobenius norms, as established by the Eckart–Young–Mirsky theorem. This low-rank approximation effectively embeds users and items into a shared latent space, enabling the estimation of missing ratings via the dot product of latent vectors. The Moore–Penrose pseudoinverse can also be conveniently expressed using the SVD, highlighting its mathematical versatility.

SVD gained prominence in the Netflix Prize competition, where it served as a core component of state-of-the-art hybrid recommendation models. Its advantages include computational efficiency on medium-scale datasets, interpretable latent factors, and competitive performance in contexts with dense interaction data. However, its linear formulation constrains its ability to capture non-linear dependencies, rendering it less expressive than neural approaches such as NCF.

Despite these limitations, SVD remains a robust baseline in recommender systems research. Its transparency and scalability make it a valuable benchmark for evaluating novel diagnostic techniques. In the present work, SVD is incorporated into the deletion diagnostics framework to demonstrate the model-agnostic applicability of the approach and to facilitate comparative analysis with NCF. This juxtaposition highlights the framework's capacity to yield consistent influence detection across fundamentally different recommendation paradigms.

\subsection{Popularity bias}

Popularity bias constitutes a pervasive challenge in recommender systems, manifesting as the disproportionate promotion of widely consumed or highly rated items at the expense of less popular, niche content. This phenomenon reduces recommendation diversity, diminishes user satisfaction, and can exacerbate inequities among content providers. In deep learning-based methods such as Neural Collaborative Filtering, popularity bias often emerges implicitly, as optimization for accuracy strongly favors recommending items that are already popular~\cite{abdollahpouri2017controlling}.

The adoption of deletion-based diagnostics provides a methodological lens for investigating popularity bias. By systematically quantifying the influence of individual items or users on recommendation outcomes, it becomes possible to determine whether popular items exert a disproportionate effect on model predictions. For example, a marked decline in performance metrics following the removal of a single highly popular item may indicate excessive model reliance on such items, serving as empirical evidence of bias.

Furthermore, the influence scores generated by deletion diagnostics provide actionable insights for mitigation strategies~\cite{steck2018calibrated}. Specifically, identifying items that dominate decision-making enables the design of reweighting schemes or regularization techniques that reduce over-reliance on popularity signals. These adjustments promote more balanced recommendation distributions, thus promoting fairness, improving user satisfaction, and facilitating the exposure of diverse content.

Moreover, popularity bias illustrates a domain in which deletion diagnostics serve not only as a diagnostic tool, but also directly inform bias-aware model design. By incorporating influence analyses into the recommendation pipeline, system developers gain a principled mechanism to align accuracy with diversity and fairness objectives.

\section{Methodological proposal}
\label{sec:methodological_proposal}

The cluster-level deletion diagnostics framework extends observation-level retraining by partitioning the user-item interaction matrix into user and item clusters that define structured blocks of interactions. Let $R \in \mathbb{R}^{n_u \times n_i}$ denote the interaction matrix, where $n_u$ is the number of users, $n_i$ is the number of items, and $R_{ui}$ represents an observed interaction, such as a rating, click, or implicit feedback.

To normalize for heterogeneous activity across users and items, diagonal matrices $D_u \in \mathbb{R}^{n_u \times n_u}$ and $D_i \in \mathbb{R}^{n_i \times n_i}$ are defined with entries
\begin{subequations}
\begin{align}
(D_u)_{uu} &= \textstyle\sum_{i=1}^{n_i} R_{ui}, \\
(D_i)_{ii} &= \textstyle\sum_{u=1}^{n_u} R_{ui}
\end{align}
\end{subequations}

The normalized interaction matrix is then computed as
\begin{equation}
\tilde{R} = D_u^{-1/2} R D_i^{-1/2}
\end{equation}

Normalization reduces the influence of high-activity users and popular items on the spectral decomposition and helps biclustering capture latent structures in the interaction patterns.

Although this normalization improves comparability across blocks of different sizes, it may bias the ranking of influence toward smaller blocks when a localized deletion produces a large per-interaction effect. Accordingly, normalized influence scores should be interpreted jointly with block cardinality, and they complement rather than replace the corresponding absolute performance changes.

Spectral biclustering is performed by computing the truncated singular value decomposition
\begin{equation}
\tilde{R} \approx U_k \Sigma_k V_k^\top
\end{equation}
\noindent where $U_k \in \mathbb{R}^{n_u \times k}$ and $V_k \in \mathbb{R}^{n_i \times k}$ contain the $k$ leading singular vectors of $\tilde{R}$. Each row of $U_k$ provides a low-dimensional embedding for a user, while each row of $V_k$ provides a low-dimensional embedding for an item. These embeddings encode latent co-variation structures that form the basis for cluster assignment.

User clusters ${\mathcal{U}_1, \dots, \mathcal{U}_K}$ are obtained by applying $k$-means clustering to the rows of $U_k$, and item clusters ${\mathcal{I}_1, \dots, \mathcal{I}_L}$ are obtained by applying $k$-means clustering to the rows of $V_k$. Each bicluster or block is defined as
\begin{equation}
\mathcal{B}_{kl} = \mathcal{U}_k \times \mathcal{I}_l
\end{equation}
\noindent which represents all interactions between a user cluster and an item cluster. This block structure provides a grouped representation of interaction patterns within the data.

For each block $\mathcal{B}_{kl}$, the corresponding interactions are removed from $R$, and the recommender model $f_\theta$ is retrained. Let $M$ be an evaluation metric, such as HR@K, NDCG@K, or RMSE. The influence of block $\mathcal{B}_{kl}$ on $M$ is computed as
\begin{equation}
\mathcal{I}(\mathcal{B}_{kl}) = \frac{ M(\theta; R) - M(\theta_{-\mathcal{B}_{kl}}; R) }{ |\mathcal{B}_{kl}| }
\end{equation}
\noindent where $\theta_{-\mathcal{B}_{kl}}$ are the parameters obtained after retraining with $\mathcal{B}_{kl}$ removed, and $|\mathcal{B}_{kl}|$ is the number of interactions in the block. Normalizing by the block size ensures comparability across clusters of different sizes.

The influence of user clusters $\mathcal{U}_k$ and item clusters $\mathcal{I}_\ell$ is computed similarly, by removing all interactions associated with the respective cluster. This allows the method to generate diagnostic scores not only at the level of individual interactions but also for groups of users, items, and their interactions.

The complete methodological procedure is summarized in Algorithm~\ref{alg:cluster_diagnostics}. The process first trains the model on the full interaction matrix, computes user and item clusters using spectral biclustering, and then sequentially evaluates the influence of user clusters, item clusters, and blocks. Retraining can leverage warm-start initialization from the full-data solution to reduce computational cost in practice. This procedure reduces the number of required retrainings from $O(n_u+n_i)$ to $O(K+L+KL)$, improving the tractability of the analysis relative to finer-grained deletion strategies.

\begin{algorithm}
\caption{Cluster-level deletion diagnostics}
\label{alg:cluster_diagnostics}
\DontPrintSemicolon
\SetAlgoLined
\SetKwInOut{Input}{Input}
\SetKwInOut{Output}{Output}
\Input{Interaction matrix $R$, recommender $f_\theta$, number of user clusters $K$, number of item clusters $L$, evaluation metric $M$.}
\Output{Influence scores $\mathcal{I}(\mathcal{U}_k)$, $\mathcal{I}(\mathcal{I}_\ell)$, $\mathcal{I}(\mathcal{B}_{kl})$.}

Train $f_\theta$ on $R$ and compute baseline $M(\theta;R)$;
Compute user and item clusters ${ \mathcal{U}_k }, { \mathcal{I}_\ell }$ using spectral biclustering;

\For{$k \gets 1$ \KwTo $K$}{
Remove interactions of $\mathcal{U}_k$ from $R$;
Retrain $f_\theta$ (warm-start);
Compute $\mathcal{I}(\mathcal{U}_k)$;
}

\For{$\ell \gets 1$ \KwTo $L$}{
Remove interactions of $\mathcal{I}_\ell$ from $R$;
Retrain $f_\theta$;
Compute $\mathcal{I}(\mathcal{I}_\ell)$;
}

\For{$k \gets 1$ \KwTo $K$}{
\For{$\ell \gets 1$ \KwTo $L$}{
Remove interactions of $\mathcal{B}_{kl}$ from $R$;
Retrain $f_\theta$;
Compute $\mathcal{I}(\mathcal{B}_{kl})$;
}
}
\end{algorithm}

The proposed framework provides useful post-hoc diagnostic signals through block-deletion analysis and can support the analysis of recommender models. By quantifying the influence of individual users, individual items, user clusters, item clusters, and their interaction blocks on model performance, the methodology can help identify segments or patterns that negatively affect accuracy, fairness, or robustness. These diagnostic outcomes may guide targeted interventions, including reweighting interactions, refining user or item representations, or retraining with augmented or filtered data to reduce over-reliance on detrimental blocks. Consequently, the post-hoc explanations can serve as a feedback mechanism for model analysis and refinement.


Spectral biclustering provides a structured way to define deletion blocks based on latent patterns in the interaction data. This allows influence scores to be analyzed at the level of grouped users, items, and their interactions. The framework is model-agnostic and can be applied to different recommender architectures, including matrix factorization, neighborhood-based methods, and neural models. Systematic evaluation of structured interaction blocks can provide insights into segment-level fragility, latent dependencies, and the ways in which different user and item segments affect model behavior, beyond what is directly visible through conventional performance metrics alone. This multi-granularity approach supports diagnostic analysis at several levels of aggregation and can inform model auditing and refinement.

A natural question concerns how the proposed block-deletion diagnostics relate to established post-hoc explainability techniques such as LIME, SHAP, and counterfactual explainers. These methods are primarily designed for feature-based predictive models and aim to attribute importance to individual input variables by fitting local surrogate models or computing marginal contributions.

In recommender systems, however, collaborative filtering models often operate on sparse user-item interaction matrices without explicit, interpretable features. In such settings, feature-attribution methods are either inapplicable or require ad-hoc feature engineering that may obscure the original data generating process. Moreover, surrogate-based explanations provide only approximate fidelity and do not capture the effects of retraining the model under counterfactual data perturbations.

The proposed framework addresses a complementary explanatory question: which segments of the interaction data appear to have a stronger effect on the model's recommendations? By deleting user--item blocks and retraining the model, the method provides a model-agnostic, retraining-based diagnostic view of how structured subsets of the interaction data affect recommendation behavior. Rather than assigning importance to individual interactions or features, explanations are expressed at the level of user segments, item categories, and their interactions, which can be useful for auditing, debugging, and bias analysis.


Importantly, these explanation modalities are not mutually exclusive. Feature-attribution methods can be used to justify individual recommendations to end users, while block-deletion diagnostics provide higher-level insights into data dependence and model behavior for developers and system designers.

\section{Experimental approach}
\label{sec:experiments}

To evaluate the effectiveness of the proposed cluster-level deletion diagnostics, a set of experiments is conducted across multiple datasets and models. The objective is to determine whether the method provides consistent and interpretable insights while remaining computationally feasible.

Experiments are conducted on two well-established benchmark families. The first consists of the MovieLens dataset, which provides explicit ratings along with rich metadata such as genres. The second includes the Amazon review datasets (Electronics), where implicit feedback is derived from user reviews or purchase histories. Collectively, these datasets cover both explicit and implicit recommendation settings, exhibiting varying levels of sparsity and scale.

Two representative recommendation algorithms are compared. The first is a Neural Collaborative Filtering model, implemented using the NeuMF architecture, which integrates a generalized matrix factorization branch with a multi-layer perceptron. The second is a matrix factorization model trained via truncated singular value decomposition, providing a simpler yet competitive baseline. Evaluating both models allows examination of whether the proposed diagnostics yield consistent results across neural and classical approaches.

To ensure full reproducibility, all experiments fix the random seed and report all hyperparameters used for biclustering, training, and evaluation. Unless stated otherwise, we use TOP-$K=10$ for ranking-based evaluation, and we apply the same train/test split across all deletion conditions.

We split interactions chronologically with a 75/25 ratio, i.e., the earliest 75\% of each user’s interactions are used for training and the most recent 25\% for testing. Given the training set $\mathcal{D}_{train}$, we construct the user-item matrix $R$ by pivoting over user and item identifiers. In the explicit-feedback setting, each entry is the mean observed rating for that $(u,i)$ pair and missing values are filled with zeros. Concretely, we compute
\[
R[u,i] = \mathrm{mean}\{r_{ui}\} \quad \text{and} \quad R[u,i]=0 \text{ if } (u,i)\notin \mathcal{D}_{train}.
\]
In the implicit-feedback setting, we binarize interactions by setting $R[u,i]=1$ whenever an interaction exists and $0$ otherwise.

User and item clusters are obtained via spectral biclustering implemented in sklearn. For each configuration $(K,L)$, we fit a new biclustering model with log normalization and a fixed seed to the interaction matrix $R$ (dense pivot table), and cluster assignments are extracted through \texttt{row\_labels\_} and \texttt{column\_labels\_}. 

For each cluster or block, the following steps are performed: interactions corresponding to the cluster or block are removed from the training set; the model is retrained or fine-tuned starting from the full-data solution; the modified model is evaluated on a fixed test set; and the change in performance relative to the baseline is computed. This protocol is applied to all user clusters, all item clusters, and all user-item blocks, thereby obtaining influence scores at three levels of granularity.

Model performance is assessed using standard ranking metrics, including Hit Rate (HR@K), Normalized Discounted Cumulative Gain (NDCG@K), and Mean Average Precision (MAP@K). For the explicit MovieLens datasets, Mean Absolute Error (MAE) is also reported. Additionally, catalog coverage and diversity are analyzed to assess the impact of deletions on the distribution of recommendations. Local explanations are obtained by examining how the rank of a recommended item for a user changes when the corresponding block is removed.

\subsection{Experiment 1: SVD on MovieLens}

The proposed block-deletion diagnostics are first evaluated on the MovieLens-100k dataset using the classical SVD algorithm implemented in the \texttt{surprise} library. The model was trained with $200$ latent factors over $30$ epochs, achieving a baseline performance of NDCG = $0.1099$ and MAP = $0.0156$ on the full dataset. 

To investigate the explanatory power of block deletions, users and items were clustered into $12$ groups each using spectral biclustering. Block-deleted models were then trained by removing all interactions within a given user–item cluster pair $(U_k, I_l)$, and their recommendations were compared against those of the full model.

\begin{figure}[t]
    \centering
    \includegraphics[width=0.9\linewidth]{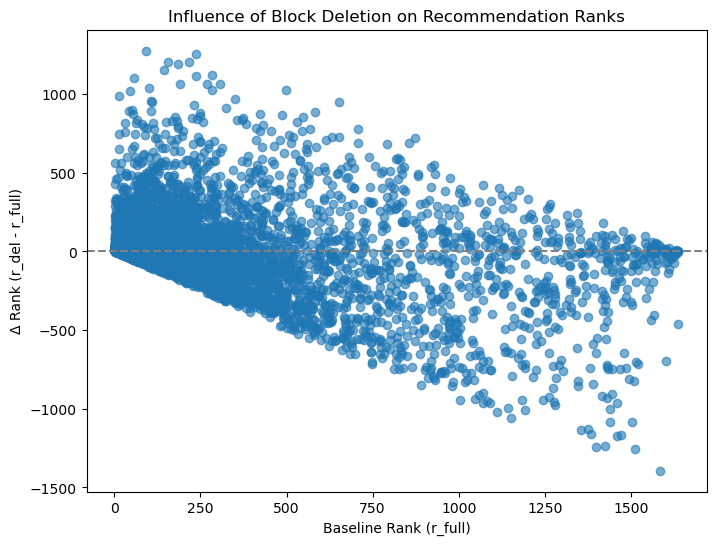}
    \caption{Influence of block deletions on recommendation ranks (SVD, MovieLens-100k).}
    \label{fig:scatter_svd_ml}
\end{figure}

Figure~\ref{fig:scatter_svd_ml} shows the relationship between baseline rank ($r_{\text{full}}$) and the change in rank ($\Delta r = r_{\text{del}} - r_{\text{full}}$) after block deletions. 
For a given user–item pair $(u,i)$, $r_{\text{full}}$ denotes the rank position of item $i$ in the recommendation list produced by the model trained on the full training set. Concretely, for each user $u$, we score all candidate items not seen in training, sort them by predicted relevance, and record the 1-based rank of item $i$.
Similarly, $r_{\text{del}}$ is computed after retraining the model with a specific user cluster, item cluster, or user–item block removed from the training data. The rank difference $\Delta r = r_{\text{del}} - r_{\text{full}}$ quantifies the counterfactual impact of that deletion.
In this scatter plot, each point in the scatter plots corresponds to a single recommendation instance $(u,i)$. The x-axis shows the baseline rank $r_{\text{full}}$, while the y-axis shows the rank change $\Delta r$ induced by deleting the corresponding explanatory block. These plots visualize the local, counterfactual effect of block deletions on individual recommendations.
It is observed that top-ranked items (low $r_{\text{full}}$) are the most sensitive to deletions, with their ranks shifting dramatically in both directions. In contrast, items with high baseline rank remain largely unaffected, suggesting that the explanatory signal concentrates on strong recommendations rather than weak ones.

\begin{figure}[t]
    \centering
    \includegraphics[width=1.0\linewidth]{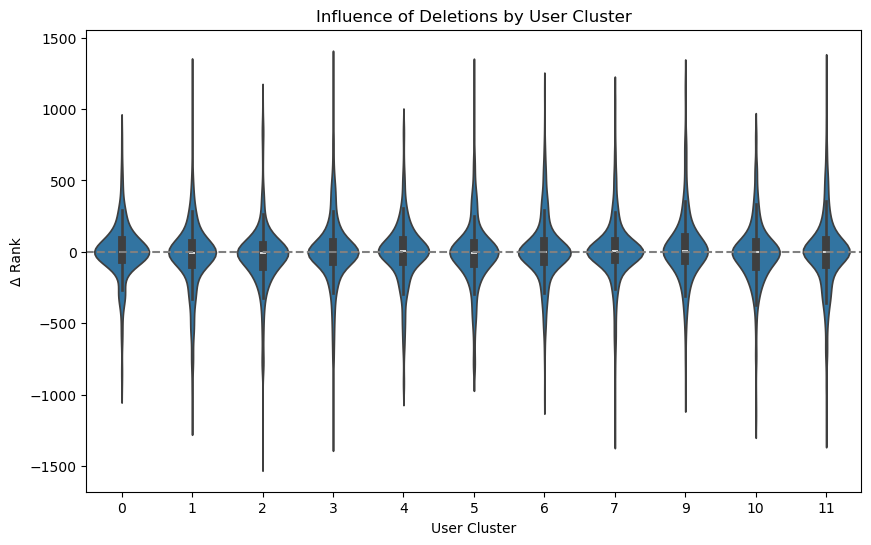}
    \caption{Distribution of rank changes $\Delta r$ across user clusters (SVD, MovieLens-100k).}
    \label{fig:violin_svd_ml}
\end{figure}

Figure~\ref{fig:violin_svd_ml} presents the distribution of $\Delta r$ across user clusters. User clusters are obtained via spectral biclustering applied to the user–item interaction matrix $R$. Each user is assigned a cluster label via the biclustering row labels. In the violin plots, rank changes $\Delta r$ are grouped by the cluster of the corresponding user, showing how sensitivity to block deletions varies across user segments. Although most clusters are centered around zero, certain clusters (e.g., $U_2$, $U_6$, $U_{10}$) exhibit much broader distributions, indicating that their recommendations are highly dependent on specific block interactions. This reveals a form of segment-level fragility: some user groups rely more heavily on particular item clusters, while others are more robust.

\begin{figure}[t]
    \centering
    \includegraphics[width=1.0\linewidth]{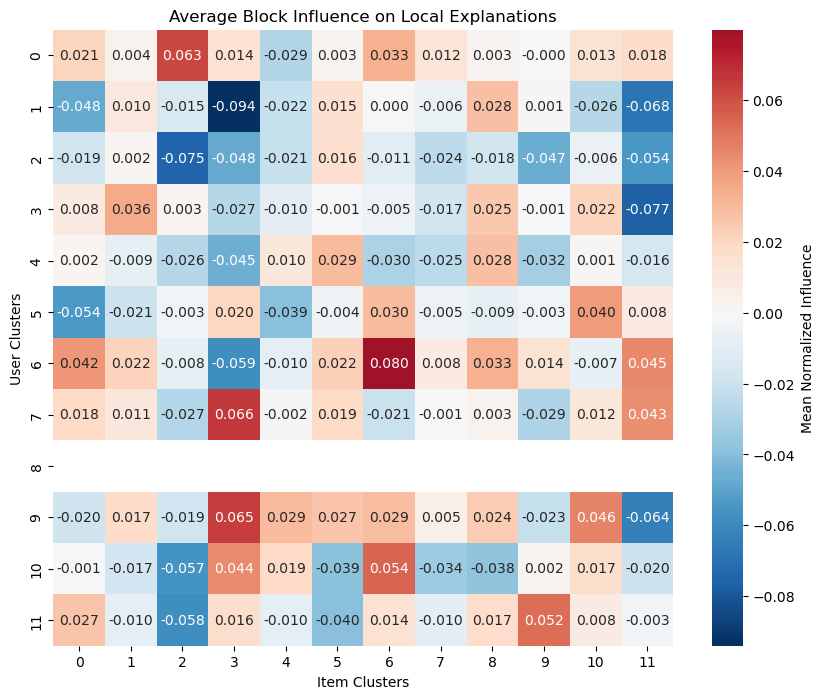}
    \caption{Average normalized influence by user-item block (SVD, MovieLens-100k). Red = supportive, blue = detrimental.}
    \label{fig:heatmap_svd_ml}
\end{figure}

Figure~\ref{fig:heatmap_svd_ml} summarizes the mean normalized influence across all user-item blocks. Each heatmap cell corresponds to a user–item block $(k,l)$, defined by user cluster $k$ and item cluster $l.$ The value shown is the average normalized influence of deleting that block, computed as the mean rank change $\Delta r$ (optionally normalized by the candidate set size) across all recommendations associated with users in cluster $k$ and items in cluster $l$. Positive values indicate supportive blocks, while negative values indicate detrimental or distracting blocks.
Most blocks contribute little explanatory signal, but a few stand out as highly influential. For instance, block $U_6 \times I_6$ shows the strongest positive contribution ($+0.08$), implying that removing these interactions severely degrades recommendation quality. Conversely, blocks such as $U_1 \times I_3$ have negative influence (–0.094), suggesting that their removal improves performance, possibly by eliminating noisy or inconsistent patterns.

Overall, these results highlight that block-deletion diagnostics provide scalable, 
counterfactual diagnostic explanations for matrix factorization models. By identifying both supportive and detrimental cluster interactions, the method extends beyond global accuracy metrics to reveal why recommendations are made and which parts of the data they rely on. This segment-level analysis is particularly useful for practitioners aiming to audit recommender systems or to detect spurious correlations between user and item groups.

\subsection{Experiment 2: NCF on Amazon}

Next, block-deletion diagnostics are applied to the Amazon dataset using a Neural Collaborative Filtering model. The model follows the NeuMF architecture, combining a generalized matrix factorization (GMF) branch with a multi-layer perceptron (MLP). The embedding dimension is set to 4 for both users and items, and the MLP consists of three hidden layers with sizes $[16,8,4]$. Models are trained for 30 epochs using a batch size of 256 and the Adam optimizer with a learning rate of $10^{-3}$. All training runs fix the random seed and reuse the same train/test split across deletion conditions. Warm-start initialization from the full-data model is employed for all retraining runs. This neural architecture provides greater flexibility compared to SVD, but also presents a higher risk of overfitting to spurious user-item correlations.

\begin{figure}[t]
    \centering
    \includegraphics[width=1.0\linewidth]{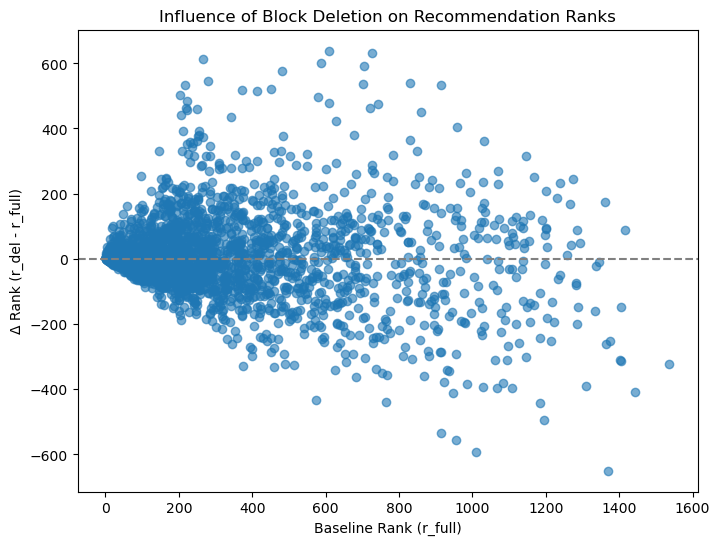}
    \caption{Influence of block deletions on recommendation ranks (NCF, Amazon).}
    \label{fig:scatter_ncf_amazon}
\end{figure}

Figure~\ref{fig:scatter_ncf_amazon} shows the effect of block deletions on recommendation ranks. As with the SVD results on MovieLens, the strongest effects appear at the head of the ranking: items with low baseline rank ($r_{\text{full}}$) are the most sensitive, exhibiting large positive and negative $\Delta r$. In contrast, poorly ranked items remain largely unaffected. This indicates that the NCF model relies heavily on specific cluster interactions to justify its top recommendations.

\begin{figure}[t]
    \centering
    \includegraphics[width=1.0\linewidth]{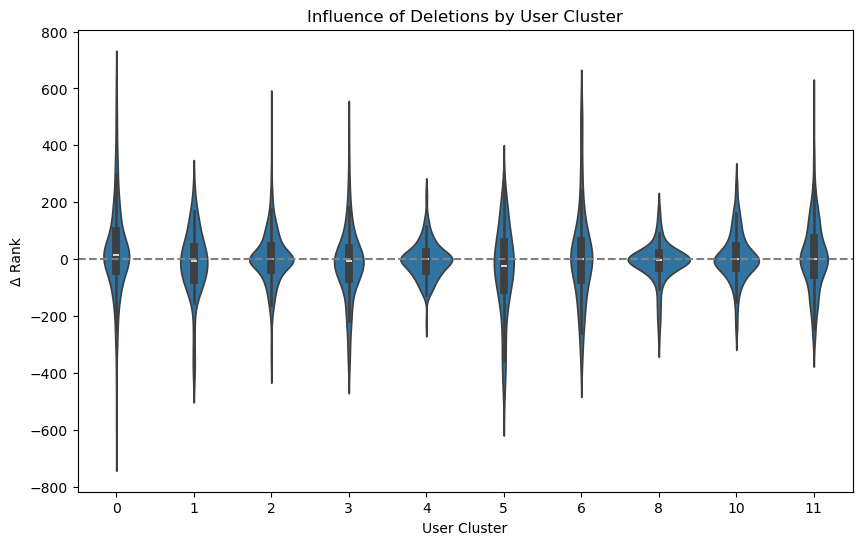}
    \caption{Distribution of rank changes $\Delta r$ across user clusters (NCF, Amazon).}
    \label{fig:violin_ncf_amazon}
\end{figure}

Figure~\ref{fig:violin_ncf_amazon} shows the distribution of rank changes grouped by user cluster. While the distributions are generally centered near zero, certain clusters (e.g., $U_0$, $U_6$, $U_{11}$) exhibit much broader spreads, with some recommendations shifting by several hundred ranks when specific blocks are removed. This indicates that the NCF model is \emph{less stable} for certain user groups, whose recommendations are disproportionately influenced by interactions with particular item clusters.

\begin{figure}[t]
    \centering
    \includegraphics[width=1.0\linewidth]{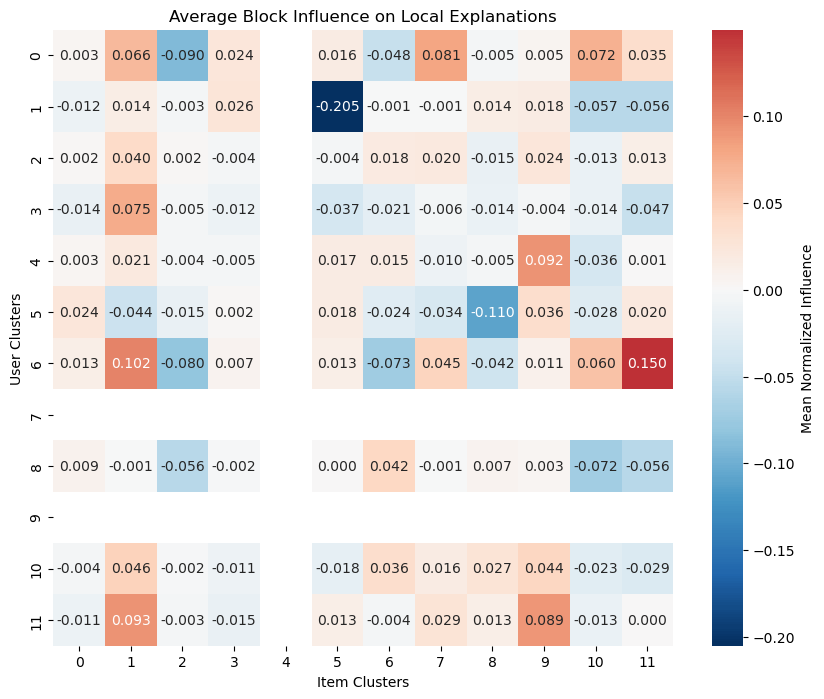}
    \caption{Average normalized influence by user-item block (NCF, Amazon). Red = supportive, blue = detrimental.}
    \label{fig:heatmap_ncf_amazon}
\end{figure}

The heatmap in Figure~\ref{fig:heatmap_ncf_amazon} reveals the mean normalized influence of each user-item block. Several blocks demonstrate strong positive influence, most notably $U_6 \times I_{11}$ ($+0.150$), indicating that the removal of this block substantially reduces recommendation quality. In contrast, other blocks (e.g., $U_1 \times I_0$ with $-0.205$) exhibit negative influence, such that their deletion leads to improved performance. These findings highlight that NCF models can amplify both beneficial and detrimental correlations in the data: some blocks provide essential collaborative signals, while others inject noise.

Block-deletion diagnostics on Amazon reveal that NCF explanations are consistent with those for SVD, but with stronger block effects. Neural recommenders appear more sensitive to specific clusters, which provides interpretable insights at the segment level but also exposes the model’s reliance on narrow patterns. This dual role-identifying supportive versus detrimental cluster interactions-demonstrates the utility of block-level diagnostics for both auditing and improving recommendation systems.

\subsection{Experiment 3: NCF on MovieLens}

\begin{figure}[t]
    \centering
    \includegraphics[width=1.0\linewidth]{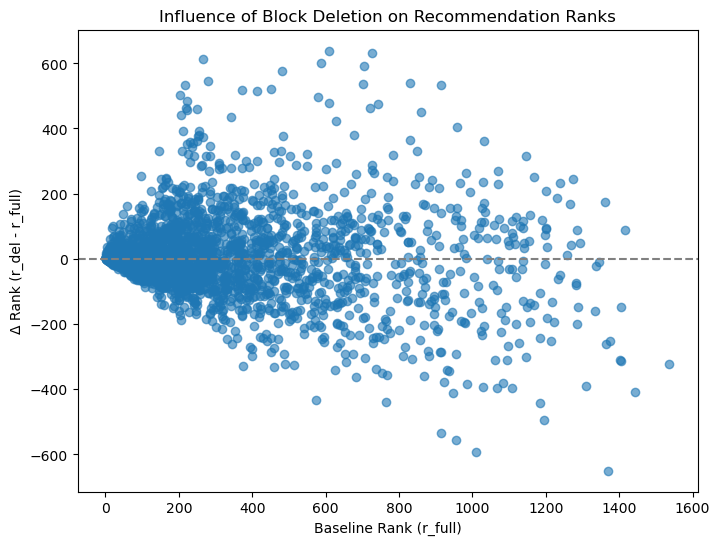}
    \caption{Influence of block deletions on recommendation ranks (NCF, MovieLens).}
    \label{fig:ncf_ml_scatter}
\end{figure}

Block-deletion diagnostics were applied to a Neural Collaborative Filtering model trained on the MovieLens dataset with the same configuration as in the second experiment. Figure~\ref{fig:ncf_ml_scatter} illustrates the relationship between baseline rank and changes in rank following block removal. Consistent with previous experiments, items with low baseline rank (i.e., strong recommendations) are the most sensitive to deletions, with shifts of several hundred positions observed in both directions. This suggests that the NCF model relies more strongly on cluster-level evidence for its top recommendations, whereas mid- and low-ranked items remain relatively stable.

\begin{figure}[t]
    \centering
    \includegraphics[width=1.0\linewidth]{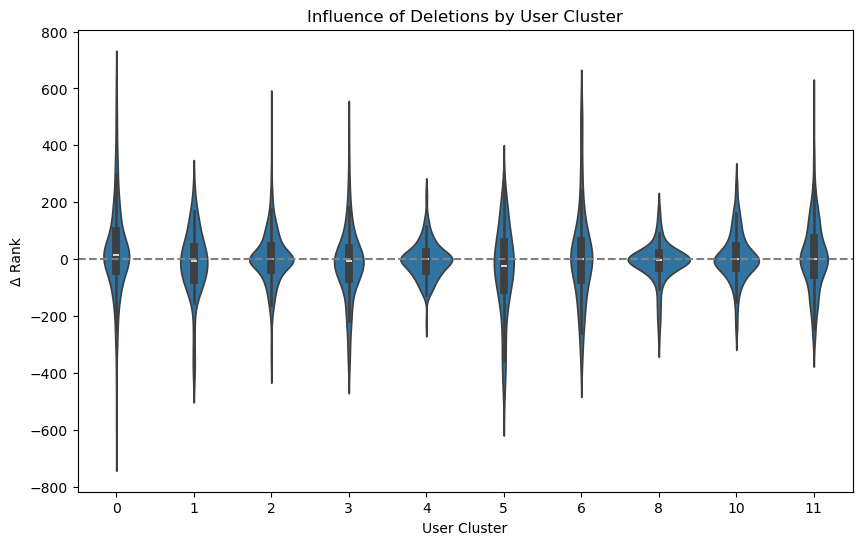}
    \caption{Distribution of rank changes $\Delta r$ across user clusters (NCF, MovieLens).}
    \label{fig:ncf_ml_violin}
\end{figure}

The violin plots in Figure~\ref{fig:ncf_ml_violin} highlight differences across user clusters. While most clusters are centered near zero, some (e.g., U0, U6, U11) exhibit much wider spreads, indicating that their recommendations are more sensitive to a small number of item clusters. Others (e.g., U2, U4) remain comparatively stable. These results suggest that the diagnostic framework can help identify user segments whose recommendations are particularly vulnerable to localized perturbations.

\begin{figure}[t]
    \centering
    \includegraphics[width=1.0\linewidth]{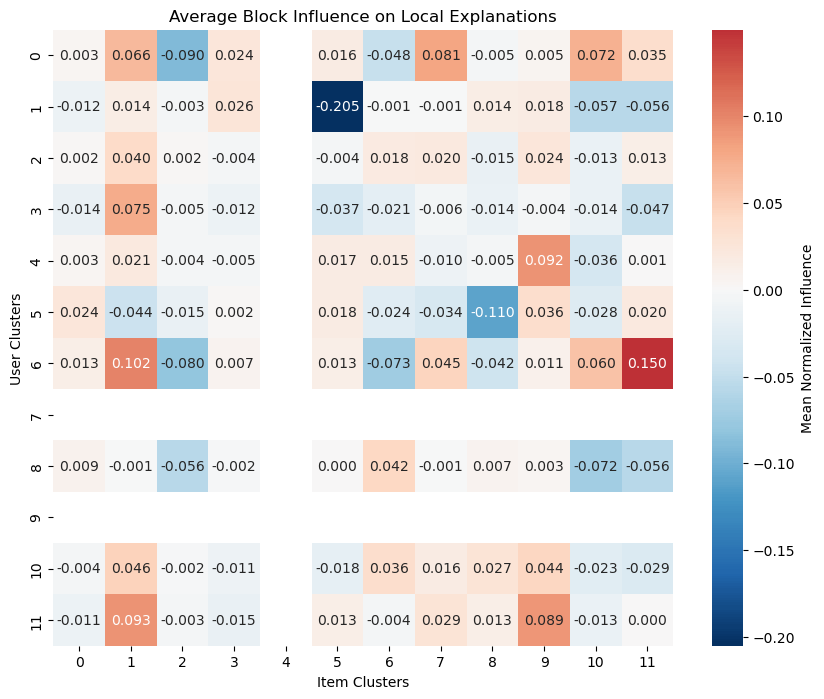}
    \caption{Average normalized influence by user-item block (NCF, MovieLens). Red = supportive, blue = detrimental.}
    \label{fig:ncf_ml_heatmap}
\end{figure}

Finally, the heatmap in Figure~\ref{fig:ncf_ml_heatmap} reveals block-level influences. Most blocks exert only minor effects, but some exhibit strongly positive or negative impacts. For example, block U6$\times$I11 has a pronounced positive influence, whereas block U1$\times$I0 appears strongly detrimental. Compared to SVD, the NCF heatmap shows greater polarization: certain blocks provide substantial supporting evidence, while others act as distractors that reduce recommendation quality. These observations suggest that NCF captures powerful collaborative signals within specific clusters but is also more susceptible to overfitting spurious correlations.

Overall, the MovieLens results reflect the general patterns observed in Amazon and SVD: strong recommendations are the most sensitive to deletions, and user segments differ in their robustness. However, the more pronounced block effects observed for NCF illustrate the potential value of block-deletion diagnostics for analyzing neural recommenders, where supportive and distractor clusters may coexist.

The proposed framework relies on an intermediate spectral biclustering step to partition users and items into explanatory units. Since the number of clusters determines both explanation granularity and computational cost, we conducted a systematic sensitivity analysis over clustering resolutions $K=L\in\{6,8,12,16\}$.

For each configuration, biclustering was recomputed from scratch and followed by the full block-deletion diagnostic pipeline. Robustness was evaluated at the level of explanations rather than clustering compactness, using correlation- and overlap-based stability measures detailed below.

Across all configurations, recommendation quality (NDCG and MAP) remains broadly similar (Table~\ref{tab:cluster_sensitivity}), suggesting that predictive performance is not strongly affected by the clustering resolution. However, the explanation-oriented stability measures are more variable. In particular, the Top-10 overlap is low across most configurations, indicating limited agreement in the identity of the most influential blocks. Stability is weaker for very coarse ($K = 6$) and very fine ($K = 16$) partitions, which is consistent with under-segmentation and over-segmentation effects. Among the tested resolutions, intermediate settings, particularly $K = L = 12$, provide the most reasonable trade-off between explanatory granularity, empirical stability, and computational cost, and are therefore used in the subsequent experiments.

\begin{table}[t]
\centering
\footnotesize
\caption{Sensitivity analysis with respect to the number of user and item clusters ($K=L$). Runtime corresponds to block-deletion retrainings. Stability is measured via average rank-correlation and top-$10$ influential block overlap across configurations.}
\label{tab:cluster_sensitivity}
\begin{tabular}{c c c c c}
\hline
$K=L$ & NDCG & MAP & Top-10 Overlap & Mean Runtime (s) \\
\hline
6  & 0.110 & 0.015 & 0.176 & 87.2 \\
8  & 0.109 & 0.016 & 0.053 & 82.4 \\
12 & 0.110 & 0.016 & 0.053 & 90.8 \\
16 & 0.109 & 0.015 & 0.000 & 75.7 \\
\hline
\end{tabular}
\end{table}

To assess the robustness of the proposed explanations, we evaluate the stability of block influence patterns across different clustering resolutions ($K=L\in\{6,8,12,16\}$). Since biclustering is recomputed independently for each value of $K$, this analysis tests whether the identified influential blocks are artifacts of a specific clustering choice or reflect stable explanatory structure.

Stability is quantified using three complementary measures: (i) Pearson correlation between block influence matrices, capturing linear agreement in influence magnitude; (ii) Spearman rank correlation, capturing agreement in relative importance; and (iii) Jaccard overlap of the top-10 most influential blocks, capturing consistency among the most explanatory units.

Table~\ref{tab:stability} reports the results. While exact block assignments naturally vary with $K$, the observed agreement across resolutions is generally limited. Pearson and Spearman correlations are weak and in some cases negative, and the Top-10 Jaccard overlap is often close to zero. These results suggest that the estimated influence patterns are sensitive to the clustering resolution, although some partial consistency is observed for intermediate resolutions. Therefore, the proposed explanations should be interpreted as resolution-dependent diagnostics rather than fully stable structures that are invariant to the choice of $K$.

\begin{table}[t]
\centering
\caption{Stability of block influence patterns across clustering resolutions.
Each row compares two clustering granularities $(K_i, K_j)$ using Pearson and
Spearman correlations between block influence matrices, as well as Jaccard overlap
of the top-10 most influential blocks.}
\label{tab:stability}
\begin{tabular}{ccccc}
\toprule
$K_i$ & $K_j$ & Pearson & Spearman & Top-10 Jaccard \\
\midrule
6  & 8  & 0.014  & -0.020 & 0.176 \\
6  & 12 & -0.143 & -0.077 & 0.053 \\
6  & 16 & -0.162 & -0.202 & 0.000 \\
8  & 12 & 0.224  & 0.186  & 0.053 \\
8  & 16 & 0.254  & 0.230  & 0.000 \\
12 & 16 & 0.077  & 0.075  & 0.053 \\
\bottomrule
\end{tabular}
\end{table}

\begin{figure}[t]
\centering
\includegraphics[width=0.85\linewidth]{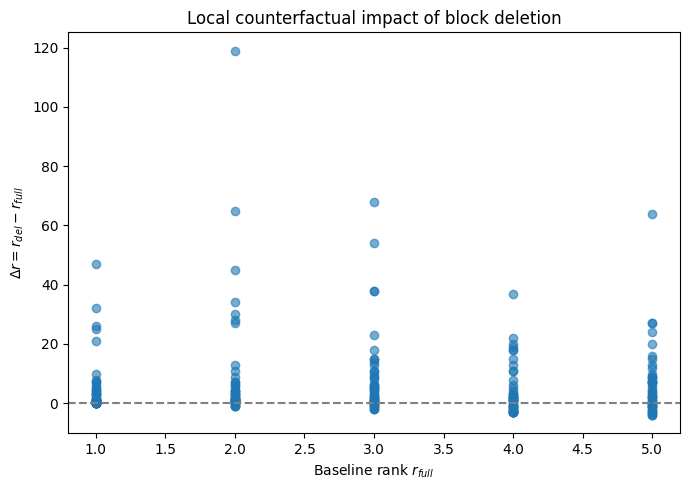}
\caption{Local counterfactual impact of block deletion on recommendation ranks. 
Each point corresponds to a user--item recommendation. The x-axis shows the baseline rank 
$r_{\text{full}}$ under the full model, while the y-axis reports the rank change 
$\Delta r = r_{\text{del}} - r_{\text{full}}$ after removing the corresponding user--item block.
Large rank shifts are concentrated among top-ranked items, suggesting that strong recommendations are more sensitive to the removal of specific blocks, while weaker recommendations remain largely unaffected.}
\label{fig:counterfactual_scatter}
\end{figure}
\begin{figure}[t]
\centering
\includegraphics[width=0.90\linewidth]{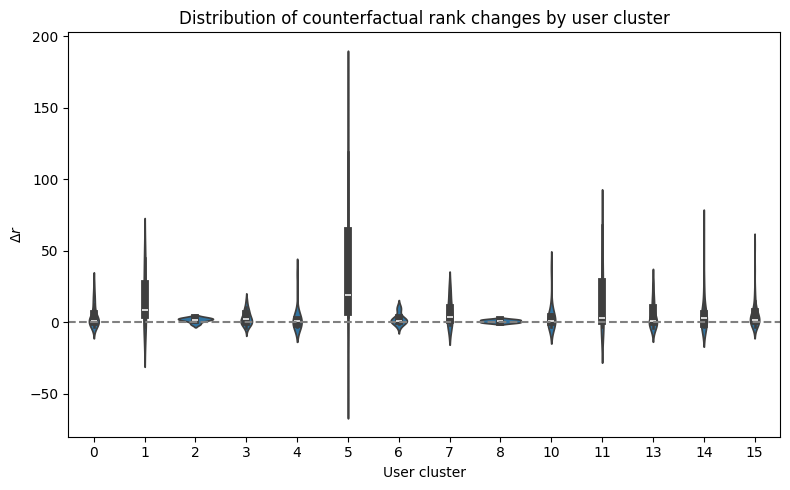}
\caption{Distribution of counterfactual rank changes $\Delta r$
grouped by user cluster. Each violin summarizes the effect
of block deletions on recommendations associated with
users from the same cluster. While most clusters are centered near zero, several exhibit wider distributions, indicating heterogeneous sensitivity to block-level perturbations. This suggests that the effects of block deletion are not uniform across user clusters, although the magnitude and stability of these effects depend on the clustering resolution.}
\label{fig:stability_violin}
\end{figure}

Beyond robustness across resolutions, we examine whether the identified blocks behave as informative counterfactual diagnostic units at the level of individual recommendations. We operationalize explanation effect through the rank change $\Delta r = r_{del} - r_{full}$ obtained after removing the corresponding user--item block. Under this definition, larger rank shifts indicate that the recommendation is more sensitive to the removed block. Figures~\ref{fig:counterfactual_scatter} and~\ref{fig:stability_violin} show that top-ranked items are often more affected by block deletions than lower-ranked ones, suggesting that some recommendations rely more strongly on specific blocks. However, this analysis should be interpreted as a rank-based diagnostic proxy rather than a complete measure of explanation quality.

\begin{figure}[t]
\centering
\includegraphics[width=0.85\linewidth]{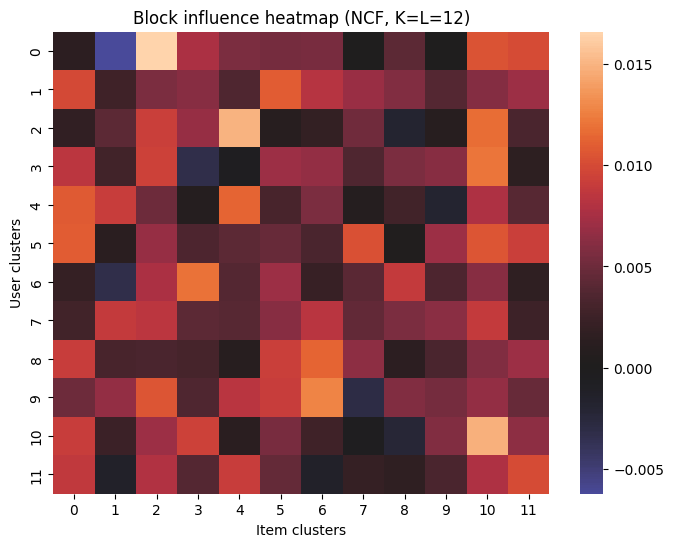}
\caption{Block density heatmap for spectral biclustering with $K=L=12$. The strong contrast between dense and sparse blocks indicates heterogeneous but coherent interaction patterns.}
\label{fig:bicluster_density}
\end{figure}

Since block-deletion explanations assume that biclusters correspond to meaningful interaction segments, we explicitly evaluate their internal quality. Figure~\ref{fig:bicluster_density} visualizes the block density matrix for $K=L=12$, revealing pronounced heterogeneity across blocks. Some user clusters interact densely with specific item clusters, while others remain sparse, indicating heterogeneous interaction patterns that are more structured than uniform or random partitions.

Quantitative validation is reported in Table~\ref{tab:bicluster_quality}. Across all tested resolutions, spectral biclustering produces denser blocks than random partitions and explains more variance in the observed ratings. This indicates that the biclustering step captures non-random structure in the interaction matrix. At the same time, these structural advantages do not by themselves imply strong stability of the downstream explanation scores, which remain sensitive to the selected resolution. Importantly, the coherence metrics remain relatively consistent across resolutions, suggesting that the biclustering step captures stable structural regularities in the interaction matrix, even though the downstream explanation scores remain only partially stable across resolutions.

\begin{table*}[t]
\centering
\footnotesize
\caption{Bicluster quality comparison between spectral biclustering and random assignments across clustering resolutions.}
\label{tab:bicluster_quality}
\begin{tabular}{ccccccc}
\hline
$K=L$ & Method & Mean density & Density std & Variance explained & Min density & Max density \\
\hline
6  & Spectral & 0.193 & 0.187 & 0.089 & 0.008 & 0.737 \\
6  & Random   & 0.050 & 0.005 & 0.004 & 0.042 & 0.059 \\
\hline
8  & Spectral & 0.211 & 0.202 & 0.100 & 0.006 & 0.825 \\
8  & Random   & 0.050 & 0.006 & 0.005 & 0.039 & 0.066 \\
\hline
12 & Spectral & 0.207 & 0.205 & 0.110 & 0.003 & 0.880 \\
12 & Random   & 0.050 & 0.008 & 0.009 & 0.033 & 0.073 \\
\hline
16 & Spectral & 0.220 & 0.214 & 0.118 & 0.001 & 0.878 \\
16 & Random   & 0.050 & 0.009 & 0.012 & 0.031 & 0.078 \\
\hline
\end{tabular}
\end{table*}

\begin{figure}[t]
\centering
\includegraphics[width=0.85\linewidth]{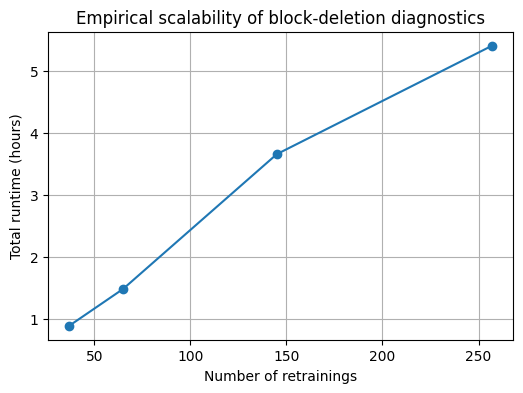}
\caption{Total wall-clock runtime of block-deletion diagnostics as a function of the clustering resolution ($K=L$).
Total runtime grows approximately linearly with the number of retrainings, while per-run cost remains stable.}
\label{fig:runtime_scalability}
\end{figure}

To assess computational cost across clustering resolutions, we measured total runtime and per-retraining runtime for $K=L\in\{6,8,12,16\}$ (Table~\ref{tab:runtime_scalability}, Figure~\ref{fig:runtime_scalability}). As expected, total runtime increases with the number of retraining runs, while per-run cost remains of similar magnitude across resolutions. This indicates that the main computational trade-off is driven by the number of diagnostic units rather than by substantial changes in the cost of an individual retraining.

\begin{table*}[t]
\centering
\footnotesize
\caption{Runtime statistics of block-deletion diagnostics for different clustering resolutions ($K=L$). 
The total number of retrainings equals $K + L + K \cdot L$.}
\label{tab:runtime_scalability}
\begin{tabular}{c c c c c c}
\hline
\textbf{$K=L$} & \textbf{Retrainings} & \textbf{Total time (s)} & \textbf{Mean (s)} & \textbf{Median (s)} & \textbf{Max (s)} \\
\hline
6  &  48  &  3225.2  &  87.2  &  77.5  &  121.2 \\
8  &  80  &  5356.9  &  82.4  &  75.3  &  115.4 \\
12 &  168 & 13172.5 &  90.8  &  79.2  &  125.0 \\
16 &  288 & 19459.9 &  75.7  &  75.5  &   93.6 \\
\hline
\end{tabular}
\end{table*}

\begin{table*}[t]
\centering
\caption{Hardware configuration and runtime decomposition of the proposed framework across clustering resolutions. The biclustering stage is executed once per configuration, whereas the retraining stage is repeated for each deleted block.}
\label{tab:runtime_split}
\begin{tabular}{lcccccc}
\toprule
$K=L$ & Biclustering (s) & Full model (s) & Retraining (s) & Total (s) & Retrainings \\
\midrule
6  & 1.90 & 260.28 & 1575.27  & 1837.45  & 36  \\
8  & 1.73 & 43.63  & 2936.51  & 2981.87  & 64  \\
12 & 1.78 & 54.28  & 7654.48  & 7710.54  & 144 \\
16 & 1.64 & 45.41  & 11406.39 & 11453.44 & 256 \\
\bottomrule
\end{tabular}
\vspace{0.5em}
\begin{tabular}{ll}
\toprule
\multicolumn{2}{l}{\textbf{Hardware configuration}} \\
\midrule
Platform & macOS-26.3.1-arm64-arm-64bit \\
Python   & 3.11.7 \\
CPU      & ARM, 10 physical / 10 logical cores \\
RAM      & 16 GB \\
GPU      & Not detected \\
\bottomrule
\end{tabular}
\end{table*}

To better characterize computational cost, we separately report the runtime of the biclustering stage and the cumulative retraining cost, together with the hardware configuration used in the experiments (Table~\ref{tab:runtime_split}). All experiments were executed on a macOS arm64 machine running Python 3.11.7, with a 10-core CPU, 16~GB RAM, and no detected GPU. Across all tested resolutions, the biclustering stage is inexpensive, requiring approximately 1.6--1.9 seconds, whereas repeated retraining dominates the total runtime. For example, at $K=L=12$, biclustering takes 1.78 seconds, while block retraining requires 7654.48 seconds over 144 deletion runs. As expected, total runtime increases with the number of deletion units, from 1837.45 seconds at $K=L=6$ to 11453.44 seconds at $K=L=16$, making explicit the trade-off between explanatory granularity and computational cost. The larger full-model time observed for $K=L=6$ is attributable to one-time initialization overhead and does not change the overall conclusion that retraining is the dominant computational bottleneck.

\begin{table*}[t]
\centering
\caption{Runtime comparison between instance-level deletion and block-level deletion on a reduced subset of the data.}
\label{tab:instance_vs_block_runtime}
\begin{tabular}{lcccccc}
\toprule
Method & Retrainings & Biclustering (s) & Full model (s) & Retraining (s) & Total (s) & Speed-up \\
\midrule
Instance-level & 830 & --   & --   & 2203.36 & 2203.36 & 1.00$\times$ \\
Block-level    & 36  & 0.52 & 2.49 & 110.49  & 113.50  & 19.41$\times$ \\
\bottomrule
\end{tabular}
\end{table*}

To provide a direct computational reference point, we compared the proposed block-level deletion framework against an instance-level deletion baseline on a reduced subset of the data (Table~\ref{tab:instance_vs_block_runtime}). On this subset, the instance-level approach required 830 retraining runs and a total runtime of 2203.36 seconds. In contrast, the block-level formulation with $K=L=6$ required only 36 retraining runs, with 0.52 seconds for biclustering, 2.49 seconds for the full-model run, and 110.49 seconds for block retraining, for a total runtime of 113.50 seconds. This corresponds to a speed-up factor of 19.41$\times$. These results confirm that the main computational advantage of the proposed framework arises from replacing a large number of fine-grained deletions with a much smaller set of structured deletion units.

The current implementation constructs a dense interaction matrix for the biclustering step, which is practical for the evaluated benchmarks but may become a memory bottleneck at much larger scales. In large sparse settings, the framework would require sparse-aware normalization and truncated SVD, or approximate alternatives such as randomized SVD, sampling-based clustering, or distributed/streaming implementations. We therefore interpret the current experiments as evidence of improved tractability on the evaluated datasets rather than as a complete demonstration of million-scale deployment.


\begin{table*}[t]
\centering
\caption{Summary of explanation-quality metrics across the 144 deleted blocks for the spectral biclustering configuration with $K=L=12$. In addition to the local rank-shift proxy $\Delta r$, we report list-level changes in recommendation quality and composition after block deletion.}
\label{tab:explanation_quality_summary}
\begin{tabular}{lccccc}
\toprule
Metric & Mean & Std. & Min. & Median & Max. \\
\midrule
$\Delta$NDCG@$K$          & 0.00388 & 0.00397 & -0.00742 & 0.00385 & 0.01488 \\
$\Delta$HR@$K$            & 0.00057 & 0.00258 & -0.00612 & 0.00050 & 0.01576 \\
$\Delta$MAP@$K$           & 0.00208 & 0.00272 & -0.00763 & 0.00208 & 0.00948 \\
Top-$K$ flip rate         & 0.42792 & 0.02435 & 0.36299  & 0.42863 & 0.49703 \\
Top-$K$ Jaccard overlap   & 0.41847 & 0.02375 & 0.35570  & 0.41750 & 0.48465 \\
Mean $\Delta r$           & -2.35974 & 11.43493 & -32.17200 & 0.75450 & 22.12200 \\
\bottomrule
\end{tabular}
\end{table*}

\begin{table*}[t]
\centering
\caption{Blocks with the strongest positive and negative effects according to $\Delta$NDCG@$K$ for the spectral biclustering configuration with $K=L=12$.}
\label{tab:top_blocks_expl_quality}
\begin{tabular}{lcccccc}
\toprule
Type & $(k,\ell)$ & $\Delta$NDCG@$K$ & $\Delta$HR@$K$ & $\Delta$MAP@$K$ & Flip rate & Jaccard \\
\midrule
Supportive   & (6,3)   & 0.01488  & 0.01576  & 0.00674  & 0.49703 & 0.35570 \\
Supportive   & (10,10) & 0.01410  & 0.00227  & 0.00948  & 0.46808 & 0.37970 \\
Supportive   & (0,2)   & 0.01340  & 0.00646  & 0.00760  & 0.48091 & 0.36833 \\
Supportive   & (2,2)   & 0.01186  & 0.00196  & 0.00778  & 0.45843 & 0.38893 \\
Supportive   & (4,4)   & 0.01045  & 0.00325  & 0.00585  & 0.44210 & 0.40610 \\
\midrule
Detrimental  & (0,1)   & -0.00742 & -0.00194 & -0.00763 & 0.42206 & 0.42315 \\
Detrimental  & (6,1)   & -0.00573 & -0.00412 & -0.00459 & 0.42651 & 0.41921 \\
Detrimental  & (3,3)   & -0.00565 & -0.00198 & -0.00401 & 0.45779 & 0.39043 \\
Detrimental  & (2,8)   & -0.00451 & -0.00612 & -0.00288 & 0.36299 & 0.48465 \\
Detrimental  & (10,7)  & -0.00413 & -0.00078 & -0.00507 & 0.43468 & 0.41240 \\
\bottomrule
\end{tabular}
\end{table*}

To complement the local rank-shift analysis, we also evaluated list-level changes after block deletion using Top-$K$ membership flips, Top-$K$ Jaccard overlap, $\Delta$NDCG@$K$, $\Delta$HR@$K$, and $\Delta$MAP@$K$ (Table~\ref{tab:explanation_quality_summary}). Across the 144 deleted blocks for the spectral biclustering configuration with $K=L=12$, block deletion consistently altered the recommendation lists, with a mean Top-$K$ flip rate of 0.4279 and a mean Top-$K$ Jaccard overlap of 0.4185. The corresponding average list-level quality changes were smaller in magnitude but non-negligible, with mean $\Delta$NDCG@$K=0.00388$, mean $\Delta$HR@$K=0.00057$, and mean $\Delta$MAP@$K=0.00208$. At the block level, the strongest supportive blocks reached $\Delta$NDCG@$K$ values up to 0.01488, whereas the most detrimental blocks reached values down to -0.00742 (Table~\ref{tab:top_blocks_expl_quality}). These results show that block deletion affects both the composition and the quality of the recommendation list. At the same time, the weak correlations between mean local rank shift $\Delta r$ and list-level changes indicate that $\Delta r$ should be interpreted as a local diagnostic proxy rather than as a direct surrogate for broader list-level recommendation disruption.


\begin{table*}[t]
\centering
\caption{Comparison of block-construction strategies on explanation-oriented metrics for $K=L=12$.}
\label{tab:baseline_comparison}
\begin{tabular}{lcccc}
\toprule
Strategy & Mean $\Delta$NDCG@$K$ & Mean $\Delta$HR@$K$ & Mean flip rate & Mean $\Delta r$ \\
\midrule
Spectral biclustering   & 0.00388 & 0.00057 & 0.42792 & -2.35974 \\
Random partitioning     & 0.00566 & 0.00130 & 0.43411 & -4.25152 \\
Activity/popularity     & 0.00373 & 0.00048 & 0.39448 & -3.05032 \\
\bottomrule
\end{tabular}
\end{table*}

To complement the structural comparison in Table~\ref{tab:bicluster_quality}, we also evaluated alternative block-construction strategies on downstream explanation-oriented metrics (Table~\ref{tab:baseline_comparison}). In particular, we compared spectral biclustering against random partitioning and an activity/popularity-based grouping baseline at $K=L=12$. The results show that the deletion effects depend on how the interaction matrix is partitioned. Spectral biclustering is competitive with these simpler baselines, but no single strategy dominates across all metrics: random partitioning yields the largest mean $\Delta$NDCG@$K$ and $\Delta$HR@$K$, whereas activity/popularity grouping yields the lowest mean Top-$K$ flip rate. We therefore interpret spectral biclustering not as uniformly superior on every list-level metric, but as a structured and non-random way of defining deletion units that remains competitive with simpler alternatives while yielding coherent partitions of the interaction matrix.

Because the single-run random baseline in Table~\ref{tab:baseline_comparison} may be sensitive to initialization, we also repeated the balanced-random block construction on a reduced subset of deletion blocks using multiple seeds. The repeated random baseline showed near-zero average effects on signed ranking-quality metrics, with mean $\Delta$NDCG@$K = 0.000105$ and mean $\Delta$MAP@$K = 0.000006$, alongside noticeable variability across runs. In particular, the mean effect of the top detrimental random blocks was unstable. These results suggest that the apparent competitiveness of a single random partition should not be overinterpreted, and that random block constructions do not provide a stable basis for diagnostic analysis. We therefore view spectral biclustering as a more structured and reproducible way of defining deletion units, while remaining competitive with simpler heuristic baselines.


To assess variability across random seeds, we repeated the block-influence analysis for the selected configuration $K=L=12$ using three different seeds (7, 13, and 29), while keeping the train/test split fixed. Across all pairwise comparisons, the resulting influence matrices were identical, yielding Pearson correlation, Spearman correlation, and Top-10 Jaccard overlap equal to 1.0, both for the overall top blocks and for the most supportive and most detrimental blocks. These results indicate that, under the current experimental setup and for a fixed train/test split, the main conclusions are stable across the tested random seeds.


\begin{table}[t]
\centering
\caption{Summary of the warm-start ablation for the NCF model, comparing warm-start and from-scratch retraining on the top three supportive and top three detrimental blocks.}
\label{tab:warmstart_ablation_summary}
\begin{tabular}{lc}
\toprule
Metric & Value \\
\midrule
Number of examined blocks & 6 \\
Same-sign rate ($\Delta$NDCG@$K$) & 0.50 \\
Same-sign rate ($\Delta$HR@$K$) & 1.00 \\
Same-sign rate ($\Delta$MAP@$K$) & 0.50 \\
Mean abs. diff. ($\Delta$NDCG@$K$) & 0.00563 \\
Mean abs. diff. ($\Delta$HR@$K$) & 0.00303 \\
Mean abs. diff. ($\Delta$MAP@$K$) & 0.00444 \\
Mean abs. diff. (mean $\Delta r$) & 6.46087 \\
Mean restored variables & 9.0 \\
Mean skipped variables & 0.0 \\
\bottomrule
\end{tabular}
\end{table}

Finally, to assess whether warm-start retraining biases the measured deletion effects, we performed a small ablation for the NCF model in which the top three supportive and top three detrimental blocks were retrained both from the full-data initialization (warm-start) and from random initialization (from scratch). Warm-start was implemented by restoring all compatible trainable variables from the full-data model and then fine-tuning on the reduced dataset; on average, all 9 trainable variables were restored and no variables were skipped. Table \ref{tab:warmstart_ablation_summary} shows mixed stability across retraining strategies. The sign of the deletion effect is preserved for all examined blocks under $\Delta$HR@$K$, but only for half of the blocks under $\Delta$NDCG@$K$ and $\Delta$MAP@$K$. The corresponding mean absolute differences are 0.00563 for $\Delta$NDCG@$K$, 0.00303 for $\Delta$HR@$K$, 0.00444 for $\Delta$MAP@$K$, and 6.46 for the mean local rank shift. These results suggest that warm-start is a useful efficiency mechanism for exploratory diagnostic analysis, but exact effect magnitudes, and in some cases even the sign of the effect under list-level ranking metrics, may depend on the retraining strategy.

\section{Conclusions}\label{sec:conclusions}

This research introduces a block-deletion diagnostic framework for explainability in recommender systems. Extending classical observation-level deletion diagnostics, the approach leverages biclustering of users and items to provide explanations at the level of user segments, item categories, and their interactions. This methodology reduces the computational cost associated with retraining relative to finer-grained deletion strategies, while providing model-agnostic counterfactual diagnostic signals.

The framework was evaluated using both matrix factorization and Neural Collaborative Filtering models across the MovieLens and Amazon datasets. First, the analysis shows that top-ranked recommendations are often more sensitive to the removal of specific user--item blocks, suggesting that strong recommendations can depend heavily on localized evidence. Second, the diagnostics reveal variations in robustness across user segments: some clusters maintain relatively stable rankings under deletion, while others exhibit larger shifts, indicating that their recommendations rely more strongly on localized signals. Third, block-level heatmaps highlight the dual role of clusters: certain blocks provide supportive collaborative evidence, whereas others act as distractors that degrade performance. This contrast is particularly visible in neural recommenders, where both strong and potentially spurious interactions may coexist, highlighting the complex structure of evidence within these models.

Taken together, these results suggest that block-deletion diagnostics provide a practical and interpretable tool for analyzing recommender systems. By combining global influence maps with local, rank-based counterfactual explanations, the methodology offers diagnostic insights beyond those directly accessible through standard performance metrics.

The contribution of this work is methodological rather than algorithmic. The proposed framework reformulates retraining-based explainability for recommender systems at the level of structured user--item blocks. By shifting the explanatory focus from individual interactions to grouped deletion units, the approach makes retraining-based counterfactual analysis more tractable on the evaluated datasets while preserving model agnosticism.

From a knowledge-based perspective, the framework provides diagnostic information about how recommendation models rely on specific segments and interaction structures. The identification of supportive, fragile, and detrimental blocks helps reveal latent dependencies and possible spurious correlations that are not directly captured by conventional performance metrics. As such, the proposed methodology can support model auditing, robustness analysis, and informed decision making in the design and maintenance of recommender systems.

Future work will extend the approach to dynamic recommendation settings, explore alternative clustering strategies, and investigate its integration into interactive explanation interfaces, thereby broadening the applicability of block-deletion diagnostics for both research and operational recommender systems.

\printcredits

\bibliographystyle{cas-model2-names}
\bibliography{cas-refs}



\end{document}